\documentclass[prd,aps,a4paper,superscriptaddress,twocolumn,nofootinbib]{revtex4}
\usepackage{graphicx}
\usepackage{color}
\usepackage{dcolumn}
\usepackage{bm}
\usepackage{slashed}
\usepackage{amsmath}
\usepackage{latexsym}
\usepackage{amssymb}
\usepackage{mathrsfs}
\usepackage{amsfonts}
\usepackage{url}
\allowdisplaybreaks
\begin{document}
\title{Amplification of gravitational wave by a Kerr black hole}

\author{Yi Gong}
\affiliation{School of Physics and Technology, Wuhan University, Wuhan, Hubei 430072, China}
\author{Zhoujian Cao
\footnote{corresponding author}} \email[Zhoujian Cao: ]{zjcao@amt.ac.cn}
\affiliation{Institute of Applied Mathematics, Academy of Mathematics and Systems Science, Chinese Academy of Sciences, Beijing 100190, China}
\affiliation{School of Fundamental Physics and Mathematical Sciences, Hangzhou Institute for Advanced Study, UCAS, Hangzhou 310024, China}
\author{Xian Chen}
\affiliation{Astronomy Department, School of Physics, Peking University, 100871 Beijing, China}
\affiliation{Kavli Institute for Astronomy and Astrophysics at Peking University, 100871 Beijing, China}

\begin{abstract}
Binary black hole may form near a supermassive black hole. The background black
hole (BH) will affect the gravitational wave (GW) generated by the binary black
hole. It is well known that the Penrose process may provide extra energy due to
the ergosphere. In the present paper we investigate the energy amplification of
the gravitational wave by a Kerr black hole background. In particular and
different from the earlier studies, we compare the energies of
the waves in the cases with and without a nearby Kerr BH. We find that
 only when the binary black hole is moving relative to the Kerr background
can the GW energy be amplified.
Otherwise, the energy will be suppressed by the background Kerr black hole.
This finding is consistent with the inequality found by Wald for Penrose
process. Taking into account realistic astrophysical scenarios, we find that
the Kerr black hole background can amplify the GW energy by at most 5 times.
\end{abstract}

\maketitle

\section{Introduction}

The Laser Interferometer Gravitational-wave Observatory (LIGO) and the Virgo
detectors have detected dozens of stellar-mass binary black holes (BBHs)
\cite{2020arXiv201014533T}. Recent astrophysical models \cite{chen19} suggest
that some of these binaries might form in the close vicinity of supermassive
black holes (SMBHs). First, BBHs can be tidally captured by a SMBH to a
distance as small as tens of gravitational radii of the SMBH without being
disrupted \cite{2018CmPhy...1...53C}. However, the event rate is low, about
four orders of magnitude below the rate inferred from the LIGO/Virgo events.
Second, the formation and merger rates of BBHs would be enhanced in the
accretion disks of active galactic nuclei (AGNs) as a result of the interaction
with the gas in the disks \cite{2012MNRAS.425..460M}. More recent studies
showed that the interaction also causes single stellar mass black holes (BHs)
to migrate towards the central SMBHs \cite{2021arXiv210109146P} and eventually
be trapped at a radius comparable to the innermost stable circular orbit (ISCO)
\cite{2021arXiv210407685P}. The accumulation of BHs at the ISCO also creates a
condition favorable to the formation and merger of BBHs. The corresponding
event rate is estimated to be $1\%$ of the LIGO/Virgo rate
\cite{2021arXiv210407685P}.

If the BBHs in the above scenarios merge, they are doing so in a curved
background induced by the nearby SMBH.  In particular, when the SMBH is
rotating, it is interesting to ask how the Kerr black hole background could
affect the gravitational waves (GW) generated by the BBH.  Especially due to
the existence of the ergosphere, ones may expect interesting energy
amplification of the gravitational wave through the Penrose process.

The energy amplification problem has been extensively studied before for particles.
Consider a particle with mass $M_{\rm F}$ and energy $E_{\rm F}$ which
decays into two daughter particles. One daughter particle falls into a negative-energy orbit,
and the other {goes out with a} mass $M_{\rm D}$ and energy $E_{\rm D}$. Wald \cite{1974ApJ...191..231W,1972ApJ...178..347B} concluded that the Penrose process does not work in this situation unless the initial particle moves with velocity larger than half of the speed of light. The energy amplification is limited by an inequality (Eq.~(4) of \cite{1974ApJ...191..231W})
\begin{align}
&\gamma_{\rm FD}\frac{E_{\rm F}}{M_{\rm F}}-\gamma_{\rm FD} v_{\rm FD}\sqrt{1+\frac{E_{\rm F}^2}{M_{\rm F}^2}}\leq\frac{E_{\rm D}}{M_{\rm D}}\nonumber\\
&\leq\gamma_{\rm FD}\frac{E_{\rm F}}{M_{\rm F}}+\gamma_{\rm FD} v_{\rm FD}\sqrt{1+\frac{E_{\rm F}^2}{M_{\rm F}^2}}
\end{align}
where $v_{\rm FD}$ is the relative velocity between the {initial} particle and the outgoing daughter particle and $\gamma_{\rm FD}=1/\sqrt{1-v_{\rm FD}^2}$.
{More recently, however, it is found that if initially there are two particles and they are
colliding near a Kerr SMBH, the total
energy could be amplified due to the Penrose process
by an arbitrary factor \cite{PhysRevLett.103.111102}.} For
this reason, Kerr black holes have been widely investigated as a particle
accelerator
\cite{PhysRevLett.104.021101,PhysRevLett.109.121101,PhysRevLett.110.011102,PhysRevLett.113.261102,PhysRevLett.114.251103,PhysRevD.93.084025,Jiang2019}.

For a BBH with a total mass of $M$, {the GW energy radiated away during the
merger is typically} $\eta M$ with $\eta\approx5\%$ \cite{PhysRevD.75.124018}. For
example, GW150914 is composed of two black holes with {initial} masses
36M${}_\odot$ and 29M${}_\odot$, and the GW carries a
total amount of energy of 3M${}_\odot$. If such a merger happens near a Kerr black
hole, how much energy will be carried away by the gravitational wave?
Especially if the final remnant black hole falls onto an orbit with a negative
energy around the Kerr black hole, what will happen? We investigate this
problem in the current paper.

We find that only when the BBH is moving relative to the Kerr BH can the GW
energy be amplified. This is similar to the finding by previous authors
\cite{1974ApJ...191..231W,1972ApJ...178..347B} for particles.
Phenomenologically, our physical set up can be viewed as a particle, constituted
by the initial binary, decaying into two other particles, corresponding to the
final black hole and the gravitational wave which moves at the speed of light.
Outside the ergosphere, astrophysical reality makes the relative speed between
the BBH and the background Kerr black hole small. In this case we find that the
Kerr BH can amplify the energy of gravitational wave by at most 1.3 times.
Inside the ergosphere, we find that the amplification is less than 5 times.

The arrangement of the paper is as following. We describe the physical setup in
the next Section. {In Sec.~\ref{sec3} and Sec.~\ref{sec4} },
we discuss the energy amplification of the
gravitational wave when the BBH merges, respectively, outside and inside the
ergosphere of the
Kerr black hole. A
summary and discussion are given in the last section. Throughout the paper we
use geometric units with $c=G=1$.

\section{Physical setup}
We assume that the total mass of the BBH is $M$, and the four velocity of the center of mass is $U^a$ respect to the background Kerr black hole. The mass of the merger remnant black hole is $M'$. Due to the kick velocity, the four velocity of the final black hole, $U'^a$,
 is different from $U^a$. Using the 3+1 decomposition of $U'^a$ relative to the observer $U^a$, we have \cite{wald84}
\begin{align}
&U'^a=\gamma(U^a+v^a),\label{eq7}\\
&\gamma\equiv-U'^aU_a,
\end{align}
where $v^a$ is perpendicular to $U^a$ which means $U^av_a=0$. Since $v^a$ is
perpendicular to $U^a$, $v^a$ is spacial respect to $U^a$ and is called the
three velocity of $U'^a$ respect to $U^a$. Consequently, $v^a$ corresponds to
the kick velocity \cite{wald84}. Since four velocity $U'^a$ satisfies
$U'^aU'_a=-1$, we have $\gamma=1/\sqrt{1-v^2}$ from the above relations. Here
$v^2=g_{ab}v^av^b\geq0$ because $v^a$ is spatial. Eq.~(\ref{eq7}) is a four
dimensional generalization of the classic formula of velocity superposition.

As a stationary spacetime, Kerr black hole admits a time-like Killing field $\xi^a\equiv\left(\frac{\partial}{\partial t}\right)^a$ where $t$ is the time coordinate of the Boyer-Lindquist coordinate \cite{wald84}. Respect to $\xi^a$ we denote
\begin{align}
&E\equiv-g_{ab}MU^a\xi^b,\\
&E'\equiv-g_{ab}M'U'^a\xi^b,\\
&E_{\rm GW}\equiv E-E'=-g_{ab}\xi^b(MU^a-M'U'^a).\label{eq1}
\end{align}
The above equations are valid at the spacetime point where the BBH
merger happens. $MU^a$ and $M'U'^a$ correspond to the four momenta
of, respectively, the initial binary and the final remnant black hole. According to
the law of conservation of momentum, $MU^a-M'U'^a$ corresponds to the four
momentum of the gravitational wave. Since the gravitational wave propagates to
null infinity along a null geodesic, $E_{\rm GW}$ defined above is a constant
along the geodesic because $\xi^b$ is
a Killing field. Moreover, $E_{\rm GW}$ corresponds to the gravitational wave energy
measured by the asymptotic observer $\left(\frac{\partial}{\partial
t}\right)^a$. These properties give our motivation to introduce the Killing
field into our analysis. The quantities $E$ and $E'$ defined above
significantly simplify our analysis in the following.

If a gravitational wave detector has a relative velocity
respect to the background black hole, the energy seen by the detector will
be $E_{\rm GW}$ multiplied by a Doppler factor. Usually the background Kerr
black hole is the supermassive black hole locating at the center of a galaxy
which moves together with the comoving frame of our Universe.
For simplicity, we assume that our GW detector also
moves with the comoving frame of our Universe. Consequently the energy
detected will be $E_{\rm GW}$ itself.

If we denote $\eta\equiv1-M'/M$, $\Pi\equiv-\xi^aU_a$ and $\zeta\equiv g_{ab}\hat{v}^a\xi^b$ with $\hat{v}^a\equiv v^a/v$, Eq.~(\ref{eq1}) results in
\begin{align}
\frac{E_{\rm GW}}{M}&=\Pi-\gamma(1-\eta)(\Pi-v\zeta).\label{eq2}
\end{align}
Here $\eta$ is the usual GW energy ratio used by current LIGO and other detectors. The value of $\gamma$ depends only on the amplitude of the kick velocity $v$. For a given Kerr black hole background, $\zeta$ depends only on the direction of the kick velocity $\hat{v}^a$. Since $\Pi$ depends on $U^a$, which is independent of the kick velocity, $\eta$, $\gamma$, $\zeta$ and $\Pi$ are mutually independent. In other words we can take $E_{\rm GW}$ as a function of four independent arguments $\eta$, $v$, $\zeta$ and $\Pi$
\begin{align}
\frac{E_{\rm GW}}{M}&=\Pi-\frac{1}{\sqrt{1-v^2}}(1-\eta)(\Pi-v\zeta).
\end{align}

Now we analyze the dependence of $E_{\rm GW}$ on $\zeta$, or equivalently, the dependence of $E_{\rm GW}$ on the direction of the kick velocity. We can decompose $\xi^a$ as
\begin{align}
&\xi^a=-\Pi U^a+\zeta\hat{v}^a+\sqrt{\Xi+\Pi^2-\zeta^2}\hat{w}^a,\\
&\Xi\equiv g_{ab}\xi^a\xi^b,
\end{align}
where $\hat{w}$ is spatial, normal, and perpendicular to $U^a$ and $v^a$, i.e., $\hat{w}^aU_a=\hat{w}^a\hat{v}_a=0$. So we have
\begin{align}
&-\sqrt{\Xi+\Pi^2}\leq\zeta\leq\sqrt{\Xi+\Pi^2}.
\end{align}
Since $\gamma(1-\eta)=\gamma M'/M>0$, Eq.~(\ref{eq2}) indicates that $E_{\rm GW}$ increases with $\zeta$. So the maximum energy of the gravitational wave is
\begin{align}
\mathcal{E}_{\rm GW1}\equiv\max_{\zeta}(E_{\rm GW}/M)=\Pi-\gamma(1-\eta)(\Pi-v\sqrt{\Xi+\Pi^2}).\label{eq4}
\end{align}
The values of $\Pi$ and $\Xi$ depend on the spacetime point where the merger happens. We note that this result is consistent with the Wald inequality given in \cite{1974ApJ...191..231W}.

We can rewrite Eq.~(\ref{eq4}) as
\begin{align}
\mathcal{E}_{\rm GW1}&=\mathcal{E}_{\rm GW0}\Pi+(1-\mathcal{E}_{\rm GW0})v\sqrt{\Xi+\Pi^2},\label{eq5}\\
\mathcal{E}_{\rm GW0}&\equiv 1-\gamma(1-\eta),
\end{align}
where $\mathcal{E}_{\rm GW0}$ corresponds to the ratio of GW energy respect to the BBH total mass when the binary is at rest and the Kerr SMBH is absent. Using Taylor expansion we have $\gamma=1+v^2/2+O(v^4)$. Up to the order of $v^2$, $\gamma\approx1$, $\mathcal{E}_{\rm GW0}\approx\eta$.

Based on the Boyer-Lindquist coordinate we can express $\Xi$ as
\begin{align}
\Xi=-1+\frac{2M_{\rm Kerr}r}{r^2+a_{\rm Kerr}^2\cos\theta}
\end{align}
Outside the ergosphere we have
\begin{align}
&r>M_{\rm Kerr}+\sqrt{M_{\rm Kerr}^2-a_{\rm Kerr}^2\cos^2\theta},\\
&-1<\Xi<0.\label{eq3}
\end{align}
Inside the ergosphere we have
\begin{align}
&M_{\rm Kerr}+\sqrt{M_{\rm Kerr}^2-a_{\rm Kerr}^2}<r\nonumber\\
&<M_{\rm Kerr}+\sqrt{M_{\rm Kerr}^2-a_{\rm Kerr}^2\cos^2\theta},\\
&0<\Xi\nonumber\\
&<\frac{a_{\rm Kerr}^2\sin^2\theta}{2M_{\rm Kerr}(M_{\rm Kerr}+\sqrt{M_{\rm Kerr}^2-a^2})-a_{\rm Kerr}^2\sin^2\theta}.
\end{align}
Note that $M_{\rm Kerr}>a_{\rm Kerr}$ due to the cosmic censorship, we have always
\begin{align}
-1<\Xi<1.
\end{align}
Consequently we have
\begin{align}
\Xi+\Pi^2\leq2\max(1,\Pi^2).
\end{align}
The maximal kick velocity is about 5000km/s \cite{PhysRevLett.107.231102}. With units $c=1$, $v_{\rm max}\approx10^{-2}$. Combining  $\mathcal{E}_{\rm GW0}\approx\eta\approx5\%$ \cite{PhysRevD.75.124018}, we have $(1-\mathcal{E}_{\rm GW0})v\lesssim0.2\mathcal{E}_{\rm GW0}$.
Given these relations, the second term in Eq.~(\ref{eq5}) becomes
\begin{align}
(1-\mathcal{E}_{\rm GW0})v\sqrt{\Xi+\Pi^2}\leq\left\{\begin{matrix}0.3\mathcal{E}_{\rm GW0},&\text{ if }\Pi^2<1\\
0.3\mathcal{E}_{\rm GW0}|\Pi|,&\text{ if }\Pi^2>1\end{matrix}\right.
\end{align}
So Eq.~(\ref{eq5}) reduces to
\begin{align}
\mathcal{E}_{\rm GW1}\leq\left\{\begin{matrix}(\Pi+0.3)\mathcal{E}_{\rm GW0},&\text{ if }\Pi^2<1\\
(\Pi+0.3|\Pi|)\mathcal{E}_{\rm GW0},&\text{ if }\Pi^2>1\end{matrix}\right.\label{eq6}
\end{align}

\begin{figure*}
\begin{tabular}{cc}
\includegraphics[width=0.5\textwidth]{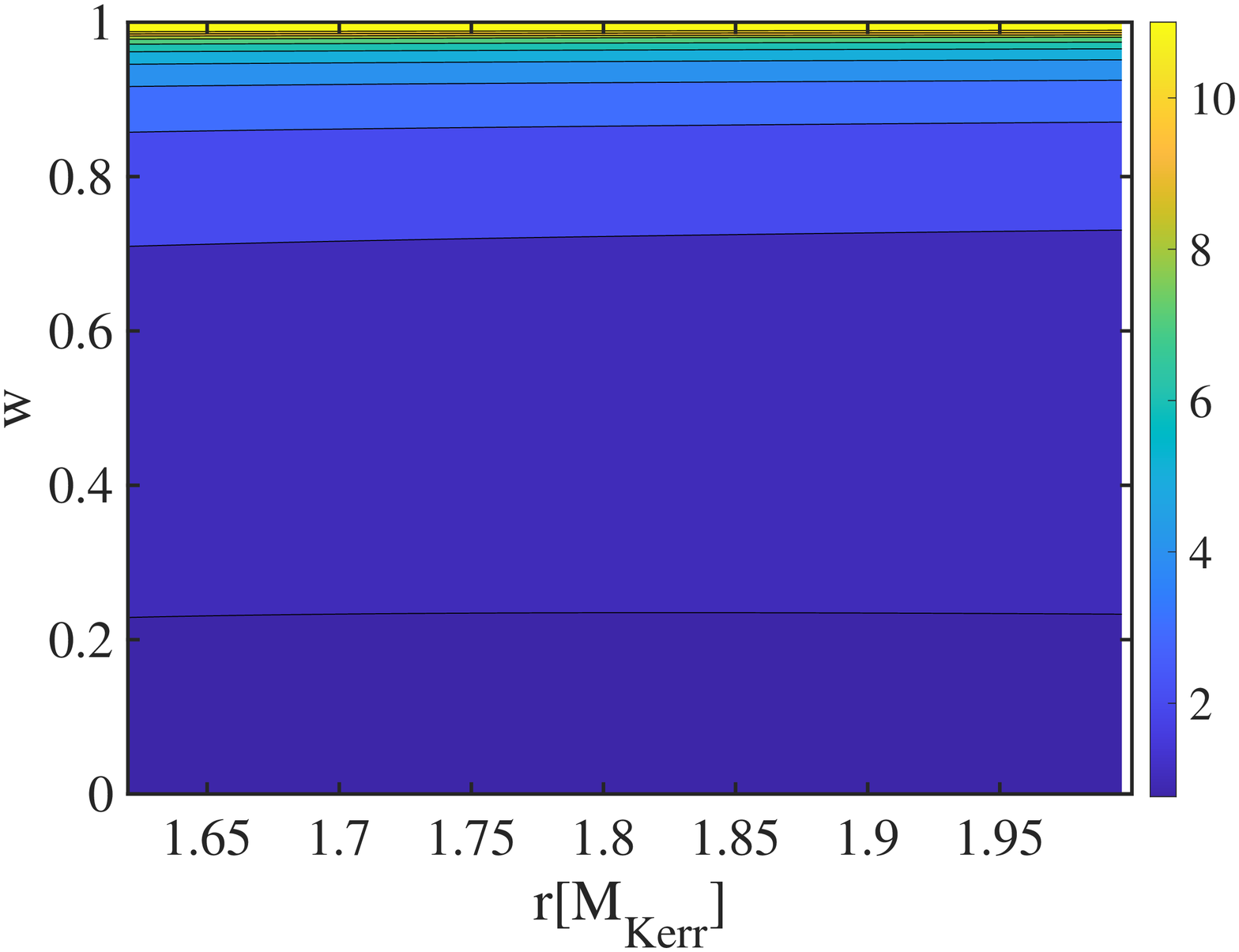}&
\includegraphics[width=0.5\textwidth]{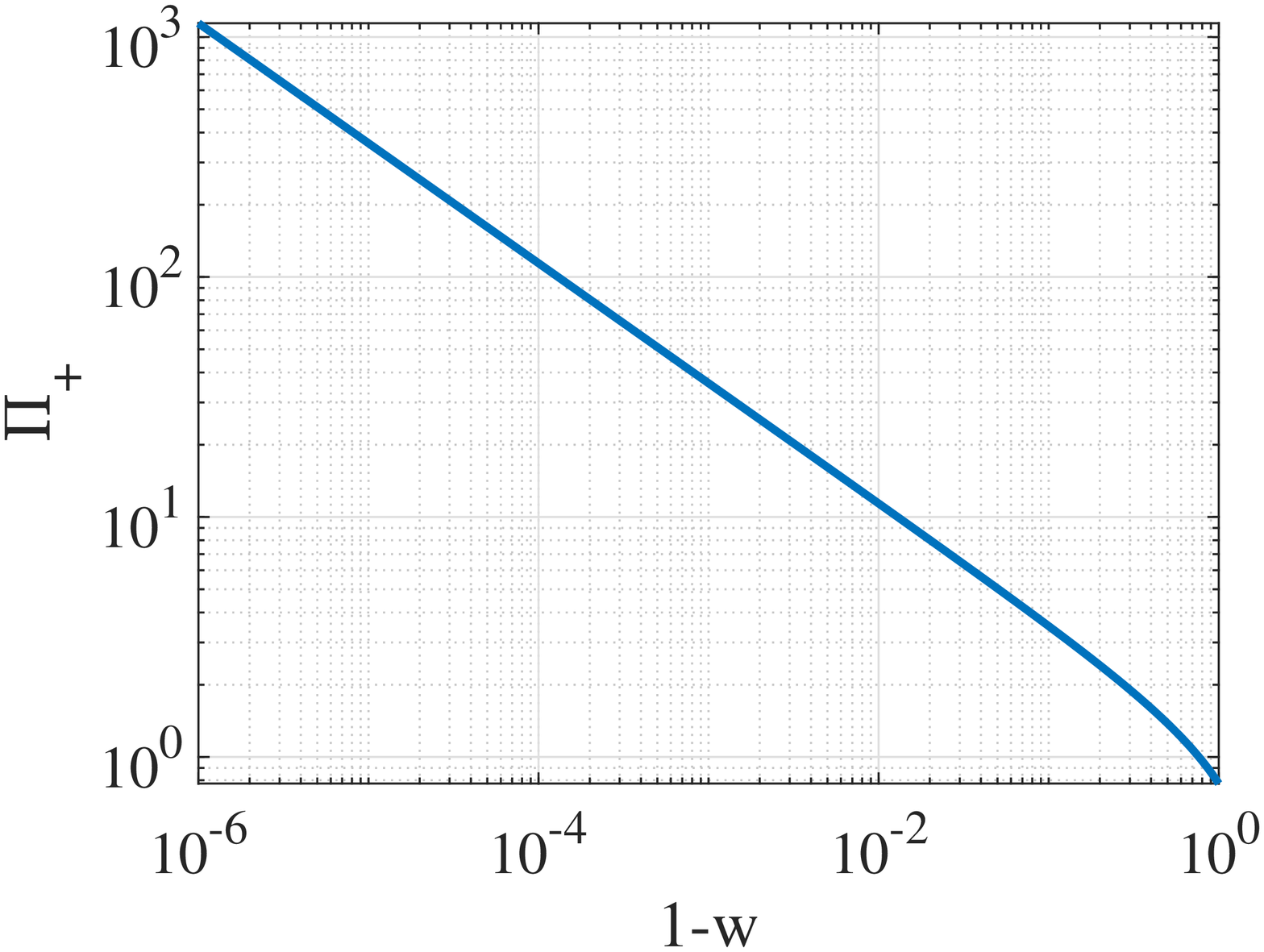}
\end{tabular}
\caption{Left: The dependence of $\Pi_+$ on $w$ and $r$. Right: $\Pi_+$
as a function of $w$ assuming that $r=1+r_{\rm ISCO}/2$ which corresponds to the center of the $r$ range in the left panel. We have set $\chi=0.98$ and $\theta=\pi/2$ in the calculation. }\label{fig1}
\end{figure*}

\section{Mergers outside the Ergosphere}\label{sec3}
As we have analyzed in the above section, $-1<\Xi<0$ outside the ergosphere. Denoting the normalized $\xi^a$ as $\hat{\xi}^a$ we have
\begin{align}
&\hat{\xi}^a=\frac{\xi^a}{|\xi|},\\
&|\xi|=\sqrt{|g_{ab}\xi^a\xi^b|}.
\end{align}

Outside the ergosphere, $\xi^a$ is time-like. If the center of mass of the BBH is at rest
with respect to the background Kerr black hole, then the four velocity of the center of mass of the BBH $U^a$ is proportional to $\xi^a$. The normalization condition of a four velocity makes $U^a=\hat{\xi}^a$ in this situation. According to the definition of $\Pi$ we have $\Pi=\sqrt{-\Xi}$ which is always less than 1 because of Eq.~(\ref{eq3}). So Eq.~(\ref{eq6}) suggests that the energy amplification factor is less than 1.3 in this situation.

If the center of mass of the BBH has a relative velocity respect to the background Kerr black hole, we have an additional Doppler factor
\begin{align}
\Pi_{\rm D}&\equiv-\hat{\xi}^aU_a,\\
\Pi&=\Pi_{\rm D}\sqrt{-\Xi}
\end{align}
where $\Pi_{\rm D}$ is the Lorentz factor between two observers whose four velocities are respectively $\hat{\xi}^a$ and $U^a$.

If $\Pi_{\rm D}<1/\sqrt{-\Xi}$, Eq.~(\ref{eq6}) suggests that the energy
amplification factor is less than 1.3. If $\Pi_{\rm D}>1/\sqrt{-\Xi}$ we have
$\mathcal{E}_{\rm GW1}<1.3\Pi\mathcal{E}_{\rm GW0}$ and the amplification factor
could be large. Since $\Pi=\Pi_{\rm D}\sqrt{-\Xi}$ corresponds to the specific
energy $E$ in Eq.~(7) of \cite{PhysRevD.75.024005}, $\Pi>1$ means that the BBH
moves along an unbound, hyperbolic orbit of the Kerr black hole background. Such kind of
binary is not realistic because astrophysical objects far from a SMBH are
moving at velocities normally much smaller than the speed of light,
which makes their orbits parabolic ($\Pi\approx1$). So we
conclude that the GW energy will be amplified by the background Kerr black hole
less than 1.3 times if the BBH merges outside the ergosphere.

\section{Mergers inside the ergosphere}\label{sec4}
Inside the ergosphere, $\xi^a$ is spatial and consequently $\Pi$ may take any
value in $(-\infty,+\infty)$ if there is no other constraint. If $\Pi<-1$,
Eq.~(\ref{eq6}) suggests that $\mathcal{E}_{\rm GW1}\leq0$. If $-1<\Pi<0$,
Eq.~(\ref{eq6}) tells us that $\mathcal{E}_{\rm GW1}\leq0.3\mathcal{E}_{\rm GW0}$.
We discuss the situations with positive $\Pi$ in the following.

If the BBH moves along a geodesic line of the background Kerr black hole, $\Pi$
corresponds to the specific energy of the geodesic line and $0<\Pi<1$ if a
bound orbit is concerned. So if the BBH forms outside the ergosphere and falls
into the ergosphere along a geodesic line,
according to Eq.~(\ref{eq6}),
the GW energy will be amplified by
the background Kerr black hole less than 1.3 times.

In the following we consider another scenario. {
It involves
an accretion disk extending to the ISCO of a Kerr SMBH, well inside the ergosphere.
Astrophysical models predict that BBHs may form in such accretion disks so that
some binaries may reside inside the ergosphere \cite{2021arXiv210407685P}.}

There is an unique future-pointing time-like normalized vector $\hat{\xi}^a_T$ which is perpendicular to $\xi^a$ and lies in the $t$-$\phi$ subspace
\begin{align}
&g_{ab}\hat{\xi}^a_T\hat{\xi}^b_T=-1,\\
&g_{ab}\hat{\xi}^a_T\xi^b=0.
\end{align}
Then $(\hat{\xi}^a_T,\hat{\xi}^a)$ form a convenient orthnormal basis for the $t$-$\phi$ subspace. In general, the matter of the accretion disk moves along a circular orbit around the Kerr black hole. In this case the four velocity of the matter $\psi^a$ locates in the $t$-$\phi$ subspace. Consequently we can decompose it as
\begin{align}
\psi^a=\psi^0\hat{\xi}^a_T+\psi^1\hat{\xi}^a,
\end{align}
then we have the relations \cite{Chandrasekhar1983}
\begin{align}
&-(\psi^0)^2+(\psi^1)^2=-1,\\
&\psi^a\xi_a=\psi^1\xi=E_c,\\
&E_c\equiv\frac{1-2u+\chi u^{3/2}}{\sqrt{1-3u+2\chi u^{3/2}}},\\
&u\equiv\frac{M_{\rm Kerr}}{r},\chi\equiv\frac{a_{\rm Kerr}}{M_{\rm Kerr}}.
\end{align}
Direct calculation results in
\begin{align}
&\psi^0=\sqrt{1+\left(\frac{E_c}{\xi}\right)^2},\\
&\psi^1=\frac{E_c}{\xi}.
\end{align}

If the BBH moves along with the accretion disk matter we have $U^a=\psi^a$ and consequently
\begin{align}
\Pi=-\psi^a\xi_a=-E_c<0.
\end{align}
Then Eq.~(\ref{eq6}) tells us $\mathcal{E}_{\rm GW1}\leq0.3\mathcal{E}_{\rm GW0}$.

However,
depending on the formation scenario of BBH, the BBH may has a relative velocity $w$
with respect to the accretion disk mater $\psi^a$. In order to get larger GW energy amplification, we need larger velocity $w$. In the viewpoint of the observer whose four velocity coincides with $\hat{\xi}^a_T$, the BBH and the accretion matter should move in opposite directions to make $w$ larger. That means the four velocity of the BBH can be expanded as
\begin{align}
&U^a=U^0\hat{\xi}^a_T+U^1\hat{\xi}^a,\\
&-(U^0)^2+(U^1)^2=-1,\\
&U^a\xi_a=U^1\xi=\Pi,\\
&\psi^aU_a=-\frac{1}{\sqrt{1-w^2}}.
\end{align}
These equations give us
\begin{align}
\Pi=\frac{\xi}{\sqrt{1-w^2}}\left[\psi^1\pm w\psi^0\right],
\end{align}
where the `+' sign corresponds to the case where the relative velocity of $U^a$
respect to $\psi^a$ is along the $\xi^a$ direction, while `-' corresponds the
$-\xi^a$ direction. When the relative velocity is along the $\xi^a$ direction the
amplification effect is stronger because the frame-dragging effect
enhances the relative motion. We
denote this bigger one $\Pi_+$. Apparently, $\Pi_+$ depends on $w$ and the
location where the BBH merger happens. As an example we set $\chi=0.98$ and
$\theta=\pi/2$ to check the dependence of $\Pi_+$ on $w$ and $r$.
The result is shown in the left panel of Fig.~\ref{fig1}. We
can see that the dependence of $\Pi_+$ on the location $r$ is negligible,
but the dependence on $w$ is more prominent.
To see more clearly the dependence on $w$, we set $r=1+r_{\rm ISCO}/2$ which corresponds to the center of the $r$ range in the left panel,
and plot
the corresponding $\Pi_+$ respective to $r$ in the right panel of the
Fig.~\ref{fig1}. We can see that $\Pi_+$ is less than 4 when the velocity $w$
is less than 0.9. Only when $w>0.99$ ($>0.9999$) can we have $\Pi_+>10$
($>100$).

\begin{figure}
\begin{tabular}{c}
\includegraphics[width=0.5\textwidth]{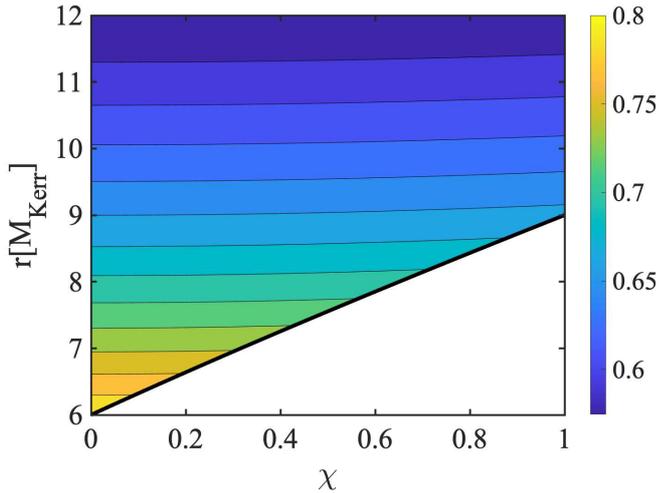}
\end{tabular}
\caption{The relative velocity between {prograde and retrograde circular orbits
as a function of the spin parameter $\chi$ and the orbit radius $r$.} The black line corresponds to the inner most stable circular orbit {for retrograde motion.}}\label{fig2}
\end{figure}

In {a realistic astrophysical scenario}, the biggest possible relative
velocity $w$ between the accretion disk and the BBH occurs when one corotates
with the spinning SMBH and the another moves on a retrograde orbit
\cite{2020ApJ...897..142S}. So we can estimate such $w$ based on the relative
velocity between the prograde circular orbit and the retrograde circular orbit.
In Fig.~\ref{fig2} we plot such relative $w$ respect to the spin parameter
$\chi$ and orbit radius $r$. The black line corresponds to the inner most
stable circular orbit of a retrograde orbit.  We find that the relative
velocity $w$ is less than $0.9$. According to Fig.~\ref{fig1} we have
$\Pi_+\lesssim4$ for real astrophysical scenario and the amplification factor
is less than about 5 according to Eq.~(\ref{eq6}).

Based on the above analysis we conclude that the Kerr black hole background will not strongly amplify the gravitational wave of BBH merger unless the binary moves extremely fast respect to the background BH. The gravitational wave energy may be decreased by the background in most astrophysical cases.

\section{Summary and discussion}
In this paper, we have shown that the Kerr black hole background
amplifies the energy of the GWs emitted from a nearby BBH only
when the binary moves fast relative to the
SMBH.
{However, it is still unclear whether the energy amplification
could leads to a detectable signature in observation because the observable in
GW astronomy is not energy, but the strain.
As has been}
pointed out before \cite{thorne88}, gravitational wave
behaves like a scalar with respect to a boost transformation of a Minkowsky space.
{In the transformation,
the gravitational wave behaves like a particle with} `boost weight zero' and `spin
weight 2'. Correspondingly, two observers with a relative velocity will
detect the same amplitude of the gravitational wave. The gravitational wave
energy difference detected by these two observers only comes from the time
shift. The presence of a SMBH near the GW source could change the above
conclusion but in the current paper
we did not
calculate the GW strain directly. How the Kerr BH background
affects GW strain is another fundamental problem. In
order to study the GW strain, a Teukolsky equation sourced by a BBH should be
used \cite{PhysRevD.103.L081501}.

If the binary forms outside the ergosphere, astrophysical model requires the
relative velocity of the binary with respect to the Kerr BH to be much smaller than
the speed of light. We have shown that the quantity $\Pi$, which
is defined near Eq.~(\ref{eq2}), is, correspondingly, less than $1$. So we conclude that
in this case the background Kerr black hole amplifies
the gravitational wave energy of the binary by at most 1.3 times.

Since the Killing vector becomes space-like inside the ergosphere, the relative
velocity between the Kerr black hole and the binary has a different
limit. If the binary forms outside the ergosphere and moves into the
ergosphere along a geodesic line and merges there, the corresponding $\Pi$ is
also limited by a bound orbit of the Kerr black hole. The GW energy
amplification is also less than 1.3 times. If the binary forms inside the
ergosphere, the most possible scenario is forming in the accretion disk which
moves along the circular orbit of the Kerr black hole. Then the relative speed
between the binary and the accretion disk, or to say the circular orbit, will
present a limit for $\Pi$. In this latter case, our analysis indicates
that the energy amplification is at most 5 times.

Ideally if two black holes fall into the ergosphere individually from outside
and meet to form a binary in the ergosphere like the scenario considered in
\cite{PhysRevLett.103.111102}, the factor $\Pi$ can approach infinity in
principle. But the possibility of forming such a binary is negligible in
galactic nuclei. In summary we conclude that the Kerr black hole background
amplifies the gravitational wave by at most a fact of $5$.

\section*{Acknowledgments}
%|--------------------------------------------------------------------|
We thank Sijie Gao and Jie Jiang for the helpful discussion. This work was supported by the NSFC (No.~11690023, No.~11721303, No.~11873022 and No.~11920101003) and by the Key Research Program of the Chinese Academy of Sciences, Grant NO. XDPB15. Z. Cao was supported by ``the Interdiscipline Research Funds of Beijing Normal University" and the Strategic Priority Research Program of the Chinese Academy of Sciences, grant No. XDB23040100.
%|--------------------------------------------------------------------|
\bibliography{refs}

\end{document}